\documentclass[pra,aps,showpacs,twocolumn]{revtex4}
\usepackage{graphicx}
\begin{document}

\newcommand{\nn}{\nonumber}
\newcommand{\ee}{\end{equation}}
\newcommand{\bea}{\begin{eqnarray}}
\newcommand{\eea}{\end{eqnarray}}
\newcommand{\wee}[2]{\mbox{$\frac{#1}{#2}$}}   % small fraction
\newcommand{\unit}[1]{\,\mbox{#1}}
\newcommand{\degree}{\mbox{$^{\circ}$}}
\newcommand{\ltish}{\raisebox{-0.4ex}{$\,\stackrel{<}{\scriptstyle\sim}$}}
\newcommand{\vs}{{\em vs\/}}
\newcommand{\bin}[2]{\left(\begin{array}{c} #1 \\ #2\end{array}\right)}
\newcommand{\p}{_{\mbox{\small{p}}}}
\newcommand{\m}{_{\mbox{\small{m}}}}
\newcommand{\tra}{\mbox{Tr}}
\newcommand{\rs}[1]{_{\mbox{\tiny{#1}}}}        % roman super or sub
\newcommand{\ru}[1]{^{\mbox{\small{#1}}}}

\title{Nondeterministic Amplifier for Two Photon Superpositions}
\author{John Jeffers}
\affiliation{Department of Physics, Scottish Universities Physics Alliance, University of Strathclyde, John Anderson Building, 107 Rottenrow, Glasgow G4 0NG, UK. john@phys.strath.ac.uk}
\begin{abstract}
We examine heralded nondeterministic noiseless amplification based on the quantum scissors device, which has been shown to increase the one-photon amplitude of a state at the expense of the vacuum-state amplitude. Here we propose using the same basic design to perform perfect amplification in a basis set of up to two photons. The device is much more efficient than several one-photon amplifiers working in tandem. When used to amplify coherent states this advantage is shown using either fidelity or in terms of probability of sucessful action, or more strikingly in a combination of the two.  
\end{abstract}
\maketitle
\section{Introduction}
It is not possible to amplify quantum states of light perfectly \cite{caves}. For a deterministic amplifier, which always works, in order to satisfy the uncertainty principle some extra noise must be added, and typically in standard quantum optical amplification this takes the form of an undetermined number of extra noise photons added to the signal. In an amplifier based on population inversion the amplified signal comes from stimulated emission and the noise source is spontaneous emission \cite{shepherd, glauber, stenholm, yamamoto, jil, qtol}. Although the extra noise photons are not always a problem for quantum devices \cite{jj,hamilton1,hamilton2} it is usually the case that useful quantum properties are destroyed or hidden by the noise.

A perfect deterministic amplifier would be able to transform coherent states by simply increasing their amplitude multiplicatively without adding noise: $| \alpha \rangle \rightarrow |g \alpha \rangle$ with $|g|>1$ as the amplitude gain. Different coherent states are not orthogonal, and so 
\bea
\nn|\langle \alpha|\beta \rangle|^2 &=& e^{-|\alpha - \beta|^2}\\ \rightarrow |\langle g \alpha|g \beta \rangle|^2 &=& e^{-g^2 |\alpha - \beta|^2} < e^{-|\alpha - \beta|^2}.
\eea
The states are transformed into a pair that are more orthogonal, which would be able to be distinguished better, and such a process is not allowed by quantum mechanics \cite{ruskai}. It is, however, possible to distinguish experimentally between nonorthogonal states without error, provided that inconclusive results are allowed \cite{ivanovic, dieks, peres, croke}. Similarly, it is possible to amplify perfectly provided that the amplifier does not always work. As long as the probability that the amplifier works is not so large that it allows nonorthogonal states to be distinguished better than the limits imposed by unambiguous error discrimination then the process is allowed.

In its usual form the quantum scissors \cite{peggscissors, babichev} (Fig. \ref{fig1}) is a postselecting device based on a pair of beam splitters, the first of which (BS1) has zero and one photon as inputs into the two input ports. One of the output ports forms the output for the device, and the other forms the input to the second beam splitter (BS2). The other input to the BS2 is any pure superposition of photon numbers, typically a coherent state. This state forms the effective input to the device. When zero and one photocounts are recorded at the outputs of BS2, corresponding to vacuum and one photon states, the output state from BS1 is the same superposition as that at the input to BS2, but with photon numbers higher than one removed - hence ``scissors". The remaining superposition of $|0\rangle$ and $|1\rangle$ has relative amplitudes which can be adjusted by altering the transmission and reflection coefficients of BS1. 

The possibility of adjusting the relative amplitudes provides a limited means of perfect amplification \cite{ralph2009,ferreyrol,ralph2010}. If the original input superposition is a weak coherent state, which has a small probability of photon numbers higher than one then the scissors device can perform the transformation 
\bea
\label{onephotonamptransform}
|\alpha \rangle \simeq |0 \rangle + \alpha |1 \rangle \rightarrow |0 \rangle + g \alpha |1 \rangle \simeq |g \alpha \rangle.
\eea
This is a valid approximate transformation provided that the state $|g \alpha \rangle$ does not have a significant two-photon amplitude $g^2\alpha^2/2$. The transformation is not deterministic, as it only works when the appropriate numbers of photocounts are recorded at the detectors. 

If the amplified two-photon component of the coherent state is significant the scissors device can still be used to perform the transformation, but the incoming state must be split using an initial set of beam splitters to provide several coherent states of smaller amplitudes, which can be recombined coherently after each has been independently amplified \cite{ralph2010}. There are two main drawbacks with this method of amplification. Firstly, each independent scissors device requires a single photon input and two detectors. Secondly the method uses several interferometer arms which must be stabilised to ensure that all of the coherent states are recombined with the same phase, so that the desired amplification is produced. 

There are other means of performing the amplification operation shown in eq. \ref{onephotonamptransform}, all of which are based on postselection. The addition and subtraction of single photons \cite{marek,fiurasek,zavatta} in its simplest noninterferometric form performs the required transformation with $g=2$. Oddly, for general gains the operation can also approximately be done if the photon addition is replaced by a noisy photon source \cite{marek,usuga}. The scheme has also been extended to amplify polarisation qubits \cite{gisin}.

The purpose of this paper is to consider nondeterministic scissors-based amplification which works ideally for superpositions of up to two photons and which does not require multiple first order interferometers. In section 2 we give an overview of the scissors amplifier as previously introduced. The next section describes an extension of this method which works in principle perfectly for superpositions of up to two photons. The fourth section compares fidelities and success probabilities for the two-photon amplifier considered here with networks of one-photon amplifiers. An overall figure of merit representing the utility of the device is introduced. The final section contains discussion and conclusions.

\section{one-photon amplifier}
\begin{figure}[h]
\centering
\includegraphics[height=4.5cm]{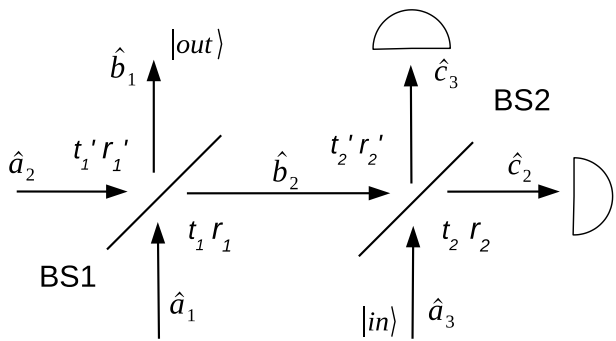}
\caption{The generalised quantum scissors device with arbitrary beam splitter coefficients, three inputs and two measured outputs.}
\label{fig1}
\end{figure}

The quantum scissors has the form shown in Fig.\ref{fig1}. It consists of a pair of beam splitters and two detectors. 
The beam splitters are denoted BS1 and BS2 and we assume that in general they can have different reflection ($r_1, r_1^\prime, r_2, r_2^\prime$) and transmission ($t_1, t_1^\prime, t_2, t_2^\prime$) coefficients for incidence from different ports. The relations between input and output operators for BS1 can be written 
\begin{eqnarray}
\label{bsrels}
\nn \hat{b}_2 &= t_1^\prime \hat{a}_2 + r_1 \hat{a}_1\\
\hat{b}_1 &= t_1 \hat{a}_1 + r_1^\prime \hat{a}_2,
\end{eqnarray} 
with similar relations for BS2. Unitarity and energy conservation require the moduli of the reflection and transmission coefficients to be equal for each beam splitter
\bea
\label{modcond}
|t^\prime|=|t|\mbox{ and }|r^\prime|=|r|,
\eea
and also imply a phase requirement
\bea
\label{phasecond}
\phi_r + \phi_r^\prime \pm \pi = \phi_t + \phi_t^\prime.
\eea
This phase requirement is a consequence of the fact that the beam splitter transformation is unitary. A less restrictive phase condition applies to lossy components. 

\begin{figure}[h]
\centering
\includegraphics[height=3.5cm]{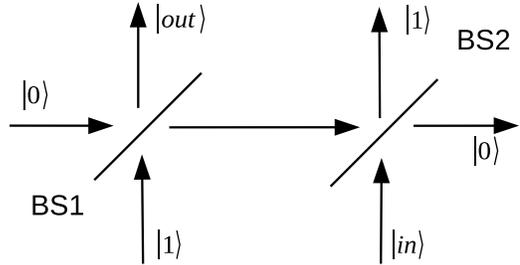}
\caption{The one photon quantum scissors device.}
\label{fig2}
\end{figure}

For a 1 photon scissors the inputs to BS1 are the vacuum and one photon states $|10 \rangle_{12}$ (Fig.\ref{fig2}), which transform to
\bea
|10 \rangle_{12} = \hat{a}_1^\dagger |00 \rangle_{12} \rightarrow \left( r_1 \hat{b}^\dagger_2+ t_1 \hat{b}_1^\dagger \right) |00 \rangle_{12}.
\eea
This forms the input in mode 2 to BS2. The other input is the coherent state, which we artificially truncate to first order in the creation operator as no more than one photon will be detected. We can therefore write the three component state as 
\bea
\nn && e^{-|\alpha|^2/2} \left( 1+ \alpha \hat{a}_3^\dagger \right) \left( r_1 \hat{b}^\dagger_2+ t_1 \hat{b}_1^\dagger \right) |000 \rangle_{123} \\
\nn &\rightarrow& e^{-|\alpha|^2/2} \left( 1 + \alpha t_2 \hat{c}_3^\dagger + \alpha r_2 \hat{c}_2^\dagger \right) \\
&\times& \left(r_1 t_2^\prime \hat{c}^\dagger_2  +  r_1 r_2^\prime \hat{c}^\dagger_3 + t_1 \hat{b}_1^\dagger \right) |000 \rangle_{123},
\eea 
where the BS2 unitary has been applied. The single photon detection in mode 3 allows the elimination of all terms proportional to $\hat{c}_2^\dagger$ and means that the mode 1 output state is 
\bea
\label{1outstate}
\nn &e^{-|\alpha|^2/2}\mbox{\rule{0mm}{0mm}} _{23}\langle 00| \hat{c}_3 \left( 1 + \alpha t_2 \hat{c}_3^\dagger \right) \left(r_1 r_2^\prime \hat{c}^\dagger_3 + t_1 \hat{b}_1^\dagger \right) |000 \rangle_{123} \\
&= e^{-|\alpha|^2/2}r_1 r_2^\prime \left( |0\rangle + g \alpha |1 \rangle \right),
\eea 
where the gain factor
\bea
g = \frac{t_1 t_2}{r_1 r_2^\prime}.
\eea
The probability that the required photocounts are obtained, and therefore that the device functions, is given by the squared modulus of this state, $e^{-|\alpha|^2} |r_1 r_2^\prime|^2 \left( 1+|g \alpha|^2 \right)$.

In the original scissors \cite{peggscissors, babichev} the gain factor was arranged to be unity, so that the device merely removed the photon number components of the coherent state which are higher than one, whilst maintaining the relative proportions of vacuum and one photon state. However it is clear that this is not the only possibility \cite{ralph2009,ferreyrol,ralph2010}. Amplification can be arranged by appropriately tuning the beam splitter reflectivities.

As the scissors device removes any two photon (or higher) state component there is an effective limit to the magnitude of the coherent state that the system will amplify. The criterion for the device to operate as a perfect amplifier is that for a particular gain $g$ any amplified two photon component would have been insignificantly small, which requires $|g^2\alpha^2|/2 \ll |g\alpha|$. If this is not the case then a better approximation to the required transformation could be made by splitting the coherent state at a series of beam splitters to reduce its amplitude, and performing the scissors amplification on each component before recombining the outputs. Such a system is heavy in terms of resources, and requires several stable interferometer arms to recombine the amplified coherent states. Furthermore, for each scissors device added the probability of success of the composite device decreases, so that for a coherent state $|\alpha \rangle $ split into $N$ equal coherent states of amplitude  $\alpha/\sqrt{N}$, and then amplified the success probability is \cite{ralph2010}
\bea
\nn P(N) &=& e^{-|\alpha|^2} |r_1 r_2^\prime|^{2N} \left( 1+\frac{|g \alpha|^2}{N} \right)^N\\
&=& \frac{e^{-|\alpha|^2}}{2^N} \frac{1}{(1+|g|^2)^N} \left( 1+\frac{|g \alpha|^2}{N} \right)^N.
\label{noutnorm}
\eea
On the second line we have assumed that BS1 is 50/50, as is typical for scissors devices, a compromise between success probability and gain, which, apart from an overall phase, is then dependent on BS2 parameters alone: $g=t_2/r_2^\prime$. 

\section{Amplification in a two-photon state space}

\begin{figure}[h]
\centering
\includegraphics[height=7.5cm]{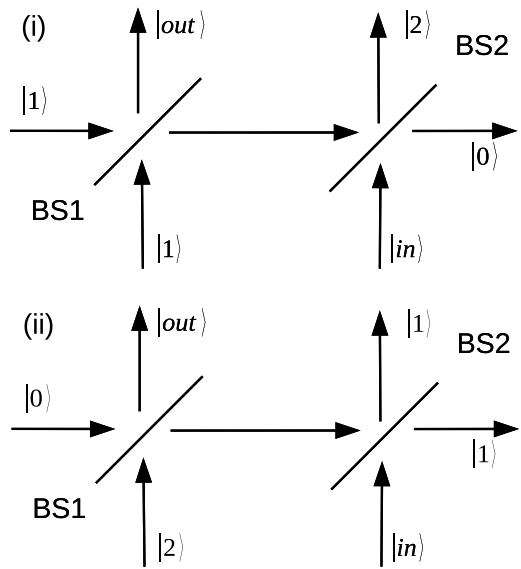}
\caption{The two photon quantum scissors device with required inputs and detector results for amplification: (i) with two single photon inputs and (ii) with vacuum and two-photon states as input.}
\label{fig3}
\end{figure}

It is natural to see if it might be possible to amplify slightly larger coherent states in a relatively simple manner. In order to do this we consider the normal scissors set-up in Fig.\ref{fig1}, but with the inputs specified in Fig.\ref{fig3}(i). For an output superposition state in mode $\hat{b}_1$ of up to two photons the input to this beam splitter should contain at least two photons. Similarly at the measurement beam splitter the input state to be amplified ought to be a superposition of $|0\rangle, |1\rangle$ and $|2\rangle$, and the measurements ought to remove a total of two photons. Such a scenario has been studied previously in the context of producing a two-photon quantum scissors \cite{koniorczyk}, where one photon enters both input ports of BS1 and one count is detected at both output ports of BS2. Other schemes have relied on the use of multiport generalisations of the quantum scissors \cite{miranovicz}. Here an amplification action on the two photon input superposition:
\bea
\label{2sup}
a_0|0\rangle +a_1|1\rangle +a_2 |2\rangle \rightarrow a_0|0\rangle +ga_1|1\rangle +g^2a_2|2\rangle
\eea
 is required for some $g$.
The input state which enters BS1 consists of a single photon in each input port, which is transformed into the output state
\bea
\label{BS1input}
\nn \hat{a}_1^\dagger \hat{a}_2^\dagger |00\rangle _{12} &\rightarrow& \\
&& \mbox{\rule{-20mm}{0mm}}\left[ t_1^\prime r_1 \hat{b}_2^{\dagger 2} + \left( t_1t_1^\prime + r_1r_1^\prime \right) \hat{b}_2^\dagger \hat{b}_1^\dagger + r_1^\prime t_1 \hat{b}_1^{\dagger 2}\right] |00\rangle _{12}.
\eea
Again the output port 2 from BS1 becomes the input port 2 for BS2. The other input port is in a coherent state, which is truncated, this time at the two photon level. This multimode state is transformed by BS2 into a three mode entangled state using the inverse relations of Eq.(\ref{bsrels}) for BS2.

It turns out that it is not possible to perform the amplifier transformation for a BS1 input of 1 photon in each input port if one count is recorded in each BS2 output port. Rather we require two photons to appear in one output port and none in the other, as shown in Fig.\ref{fig3}(a). If we assume that the two photons are detected in mode 3 this leads to the mode 1 output state
\bea
\nn &&\sqrt{2} e^{-|\alpha|^2/2} \\ 
\nn &\times&\left[ t_1^\prime r_1 r_2^{\prime 2} |0\rangle + \alpha \left( t_1t_1^\prime + r_1r_1^\prime \right) t_2 r_2^\prime |1 \rangle + \frac{\alpha^2}{\sqrt{2}} r_1^\prime t_1 t_2^2 |2\rangle \right] \\
\nn &=& \sqrt{2} e^{-|\alpha|^2/2} t_1^\prime r_1 r_2^{\prime 2} \\
&\times& \left[ |0\rangle + \alpha \frac{t_1t_1^\prime + r_1r_1^\prime}{t_1^\prime r_1} g |1 \rangle + \frac{\alpha^2}{\sqrt{2}} \frac{r_1^\prime t_1}{t_1^\prime r_1} g^2 |2\rangle \right],
\label{2outstate}
\eea
with
\bea
g=t_2/r_2^\prime.
\eea
There are two criteria for amplification which can be imposed: either we can demand pure state amplification in which the one and two photon gains are identical, or we can be less restrictive, and merely demand that the modulus of the two gains be equal. 

\subsection{Pure amplification}
The identification of gain $g$ with the BS2 parameters depends on the following condition being satisfied for BS1:
\bea
\frac{r_1^\prime t_1}{r_1t_1^\prime} = \left( \frac{t_1t_1^\prime + r_1r_1^\prime}{r_1t_1^\prime} \right)^2,
\eea
which becomes
\bea
\label{gaincond}
\nn |r_1r_1^\prime|^2 &+&|r_1r_1^\prime||t_1t_1^\prime| e^{i(\Phi_t+\Phi_r)}\\
 &+&|t_1t_1^\prime|^2 e^{2i(\Phi_t+\Phi_r)} = 0,
\eea
where $\Phi_r = \phi_r+\phi_r^\prime$ and 
$\Phi_t = \phi_t+\phi_t^\prime$. The phase condition, Eq. (\ref{phasecond}), allows this to be rewritten as 
\bea
\label{rtcondition}
\left|\frac{r_1r_1^\prime}{t_1t_1^\prime}\right|^2 - \left|\frac{r_1r_1^\prime}{t_1t_1^\prime}\right| + 1 = 0,
\eea
which can not be satisfied for real $|r_1r_1^\prime|/|t_1t_1^\prime|$. Thus there is no possible solution for BS1 which allows perfect amplification. The amplification condition (\ref{gaincond}) requires $|t_1t_1^\prime|=|r_1r_1^\prime|$, and $\Phi_t-\Phi_r = \pm 2\pi/3$. This analysis is general enough that it precludes even generalised beam splitters based on interferometers from satisfying the condition.

Beam splitters with less restrictive phase relations are allowed, but only if they are lossy \cite{barnettlossbs,jjlossbs}. As the loss of the beam splitter is increased the range of allowed phases in Eq.\ref{phasecond} around $\pi$ increases, and once the beam splitter reaches a loss level of 50\% any phase is allowed. It is shown in the appendix that the minimum loss level required to satisfy the condition (\ref{gaincond}) is 1/3. Thus BS1 can be no more than 33/33 in order for the scissors to act as a two-photon amplifier.

Of course it is not strictly necessary to have a real lossy beam splitter, and it is probably not desirable in any case, as such a component is restrictive. The lossy beam splitter can be replaced by lossless components, either a tritter \cite{multiport,tritter} in which one of the outputs is ignored, or the lossy beam splitter noise equivalent circuit \cite{barnettlossbs}. These allow the amount of loss to be controlled, and further provide the required phase freedom. It is also possible to control the amount of loss using polarising beam splitters. 

The system described here is not the only possibility. The same amplification transformations can be made if BS1 is lossless and BS2 is lossy, provided that the inputs to BS1 are $|2 \rangle$ and $|0 \rangle$ as in Fig.\ref{fig3}(ii). In this case each detector at BS2 must detect one count. However, it is more difficult to prepare a pure two-photon state at the input - a difficulty which is potentially offset by the increased detection simplicity for this system; photon number discriminating detectors are not required.

\subsection{Two-photon sign-shift amplifier}

The condition on the BS1 parameters (Eq.(\ref{rtcondition})) ensures that pure amplification is not possible for a lossless BS1. However, if the requirements are relaxed a little so that only the magnitudes of the one and two photon gains are required to be equal, 
\bea
\left|\frac{r_1^\prime t_1}{r_1t_1^\prime}\right| = 1 = \left| \frac{t_1t_1^\prime + r_1r_1^\prime}{r_1t_1^\prime} \right|^2,
\eea
so that
\bea
\left| |t_1|^2 e^{i\Phi_t} + |r_1|^2 e^{i\Phi_r}\right|^2 = \left| r_1t_1^\prime \right|^2.
\eea
The beam splitter amplitude and phase conditions allow this to be written as 
\bea
5|t_1|^4 -5|t_1|^2 +1 = 0,
\eea
which has the solutions
\bea
\label{ansst}
|t_1|^2 = \frac{1}{2} \pm \frac{\sqrt{5}}{10} \simeq 0.72, 0.28.
\eea
It is relatively straightforward to calculate the phases of the gains experienced by the one and two photon components, and it is found that the device makes the transformation 
\bea
a_0|0\rangle +a_1|1\rangle +a_2 |2\rangle \rightarrow a_0|0\rangle +ga_1|1\rangle - g^2a_2|2\rangle,
\eea
where 
\bea
g=\frac{t_2}{r_2^\prime}e^{(\phi_{t1}-\phi_{r1})}.
\eea
The lossless scissors device therefore can make the required amplifier transformation, but with a sign change to the two-photon component of the state. There are situations in optical quantum information processing where such a sign change is required. The nonlinear sign shift gate \cite{nss} performs such a sign change on the two-photon component of a state without amplification. If the amplifier sign change is not desirable the nonlinear sign shift gate could of course be used to correct it. Another possibility might be to pass the state through two identical amplifiers, in which case the nonlinear sign changes would cancel. 

The remarks at the end of the last subsection regarding interchanging the roles of BS1 and BS2 also apply here. The same transformation is made with a two photon state as input to BS1, for BS2 transmission and reflection coefficients which satisfy Eq. (\ref{ansst}), when one count is recorded at each output port of BS2.

\section{Fidelity and success probability}
\begin{figure}[h]
\centering
\includegraphics[height=5cm]{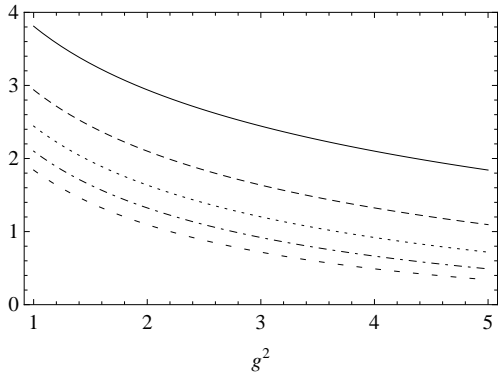}
\caption{Log of the reciprocal fidelity defect, $-\mbox{log}_{10}(1-F)$, as a function of gain for various coherent state amplitudes: $|\alpha|^2=0.1$ (Full line), $|\alpha|^2=0.2$ (dashed), $|\alpha|^2=0.3$ (dotted), $|\alpha|^2=0.4$ (dot-dashed), $|\alpha|^2=0.5$ (sparsely dashed). An ordinate value of 1 corresponds to a fidelity of 0.9, 2 to 0.99 etc.}
\label{changealpha}
\end{figure}

\begin{figure}[h]
\centering
\includegraphics[height=5cm]{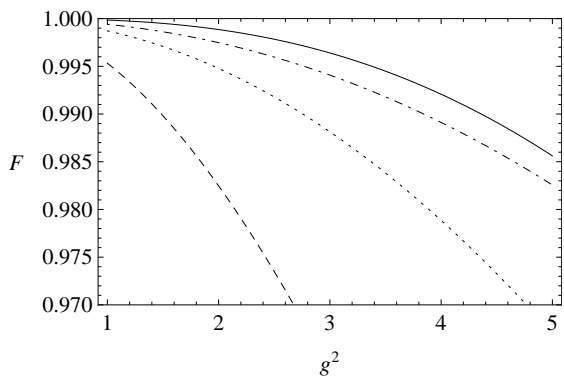}
\caption{Fidelity $F$ as a function of gain: comparison of 2-photon amplifier with N 1-photon amplifiers and for $|\alpha|^2 = 0.1$. The lines are: 2-photon amplifier (Full line), N=1 (dashed), N=2 (dotted), N=3 (dot-dashed).}
\label{alpha0.1changen}
\end{figure}

\begin{figure}[h]
\centering
\includegraphics[height=5cm]{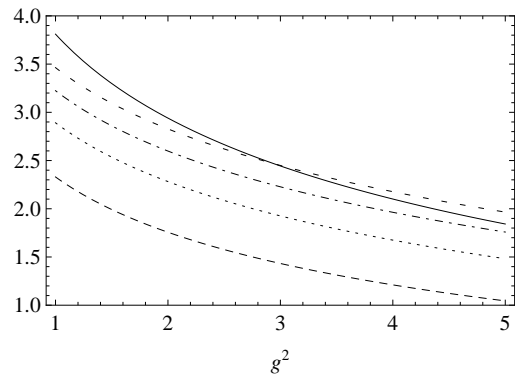}
\caption{$-\mbox{log}_{10}(1-F)$, as a function of gain: comparison of 2-photon amplifier with N 1-photon amplifiers for $|\alpha|^2 = 0.1$. The lines are: 2-photon amplifier (Full line), N=1 (dashed), N=2 (dotted), N=3 (dot-dashed), N=4 (sparsely dashed).}
\label{logalpha0.1changen}
\end{figure}

\begin{figure}[h]
\centering
\includegraphics[height=5cm]{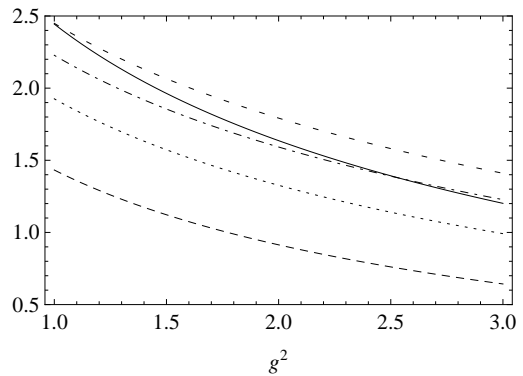}
\caption{Same as Fig. \ref{logalpha0.1changen} but with $\alpha = 0.3$.}
\label{logalpha0.3changen}
\end{figure}

The amplifier described above works perfectly for any superposition state which contains no more than two photons, but for states with nonzero amplitudes for higher photon numbers the fidelity will be less than unity. We show this effect in Figs.\ref{changealpha}-\ref{logalpha0.3changen}, which detail functions of the the squared overlap (fidelity) between the amplified output and the state $|g\alpha \rangle$ for a range of intensity gains $g^2$ and input coherent state amplitudes $\alpha$. 
In Fig.\ref{changealpha} the effect of the input state coherent amplitude on the reciprocal of the fidelity defect of the output with the coherent state $|g\alpha \rangle$ is shown; for a supposedly perfect amplifier this function provides an appropriate measure of quality - it is important to minimise the probability that the state is incorrect. The defect is tiny at lower gains, e.g. for a twofold gain the defects for $|\alpha|^2=0.1, 0.2$ are smaller than 0.001 and 0.01 respectively. Even for a fivefold gain the increases from these values are less than one order of magnitude. However, for higher coherent state amplitudes and gains the fidelity defect is larger. 

In Figs.\ref{alpha0.1changen}-\ref{logalpha0.3changen}, for comparison we show plots for the output of $N$ one-photon amplifiers. It is clear that there is considerable advantage in using only one two-photon amplifier in most cases. The set of one-photon amplifiers produces a state with higher fidelity only when two criteria are fulfilled. Firstly, that $|g\alpha \rangle$ has relevant three photon and higher amplitudes, and secondly, that $N$ is large enough to provide a significant fraction of the required amplitude. As we shall see, however, the main disadvantage to increasing the number of one-photon amplifiers is a reduced probability that the device will function. In any case, however, for $|\alpha|^2=0.1$ Fig.\ref{alpha0.1changen} shows that it is much better to use a two-photon amplifier than two or three one-photon amplifiers for reasonable gains. Even four one-photon amplifiers do not produce a stare with a higher fidelity than the two-photon amplifier for gains below about three, as shown in Fig.\ref{logalpha0.1changen}. Higher input coherent states decrease the overlap of the two-photon amplifier with the output coherent state more than they do for higher numbers of one-photon amplifiers due to the fact that the two-photon amplifier produces a state with a lower maximum photon number. In Fig.\ref{logalpha0.3changen}, which is for $|\alpha|^2=0.3$, we can see that four one-photon amplifiers do produce a state with a higher fidelity than the two-photon amplifier, but three one-photon amplifiers do so only if the gain is larger than about 2.5. 

\begin{figure}[h]
\centering
\includegraphics[height=5cm]{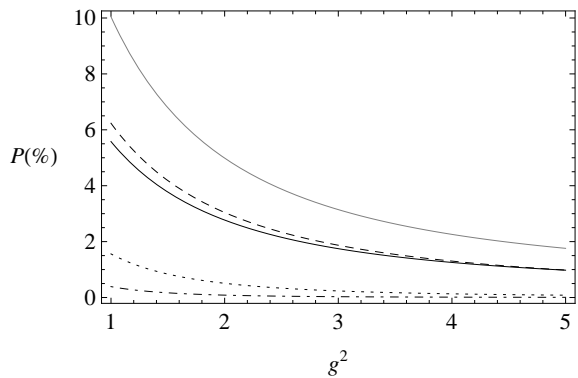}
\caption{Amplifier success probability as a function of gain for $|\alpha|^2=0.1$. The lines are: 2-photon amplifier (Full line), N=2 (dashed), N=3 (dotted), N=4 (dot-dashed) and sign-shift amplifier (grey).}
\label{palpha0.1}
\end{figure}

\begin{figure}[h]
\centering
\includegraphics[height=5cm]{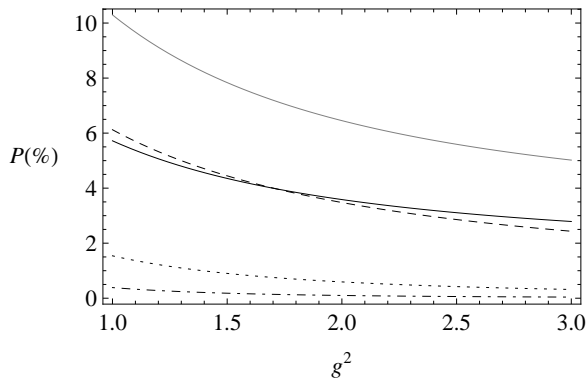}
\caption{As for Fig. \ref{palpha0.1}, but with $|\alpha|^2=0.3$.}
\label{palpha0.3}
\end{figure}

Overlap with the output state is not the only criterion for a useful device. It is also pertinent to consider the probability that the device performs the amplification. The probability that the two-photon amplifier works is given by the squared normalisation factor of the output state from Eq. (\ref{2outstate}). This is 
\bea
\nn P &=& 2 e^{-|\alpha|^2} |t_1^\prime|^2 |r_1|^2 |r_2^\prime|^4 \left( 1+|g \alpha|^2 + \frac{|g \alpha|^4}{2} \right)\\
&=& \frac{2}{9} \frac{1}{(1+|g|^2)^2} \left( 1+|g \alpha|^2 + \frac{|g \alpha|^4}{2} \right),
\eea where in the second line we have set BS1 to be symmetric, $|t_1^\prime|^2 =|r_1|^2 = 1/3$. For the two-photon sign-shift amplifier the factor of $2/9$ becomes 0.4, but the form of the probability is unchanged. We plot this in Figs. \ref{palpha0.1} and \ref{palpha0.3} against gain together with the corresponding probabilities for $N$ one-photon amplifiers from Eq. (\ref{noutnorm}). 
The two-photon amplifier success probability is of the same order as the one-photon amplifier probability for $N=2$, but is much larger than that for $N=3$ and higher (the difference will be even greater when the probability of producing the extra input photon is taken into account). The picture is not dramatically changed by increasing the input coherent state amplitude. The two-photon amplifier is slightly more likely to function at higher gains than the pair of one-photon amplifiers. The other point to note is that the two-photon sign-shift amplifier is significantly more likely to work than either the two-photon amplifier or two one-photon amplifiers. 

We can combine the fidelity defect and the success probability into an appropriate overall figure of merit, $U=P/(1-F)$ values of which represent the utility of a device for producing the desired quantum state
\begin{figure}[h]
\centering
\includegraphics[height=5cm]{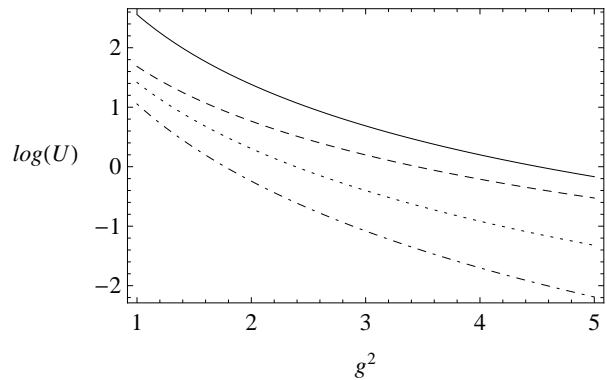}
\caption{$log_{10}$ of the utility function as a function of gain for $|\alpha|^2=0.1$. The lines are: 2-photon amplifier (Full line), N=2 (dashed), N=3 dotted and N=4 (dot-dashed).}
\label{utilityplot}
\end{figure}
It is clear from Fig. \ref{utilityplot} that the two-photon amplifier is much better than any of the other combinations of single photon amplifier according to this criterion. For low gains the two-photon amplifier is better than $N=2$ by about an order of magnitude and for gains up to about fivefold the two-photon amplifier is better by a about a factor of two. 

\section{Conclusions and discussion}

Some optical quantum information protocols are performed in the two photon basis, and require a significant two-photon amplitude, something which can be difficult to achieve from coherent states without exciting the three or higher photon number components of the state. In this paper it has been shown that it is possible to perform perfect amplification of a superposition of up to two photons using beam splitters in a quantum scissors configuration. There are two types of possible amplification. If amplification in which the two photon and one photon gains are identical is required the transformation can be performed with one lossy and one lossless beam splitter. The lossy beam splitter must on average absorb at least one third of the light which falls upon it. Such a lossy device can be straightforwardly mimicked by higher order multiports. 

The second type of amplification which is possible is that in which the two-photon component of the state receives a sign change. Such a sign change is sometimes required in optical quantum information protocols. When it is not required it can be corrected. This type of amplification can be done with two standard lossless beam splitters. 

If the amplifier is used on coherent states it will not amplify perfectly. The fidelity of the output state obtainable in this case using the two
photon amplifier, for reasonable gains and for the coherent amplitudes
typically used in quantum information, is much better than that obtainable
even with a network of three one-photon amplifiers. Also, the probability
that the two-photon amplifier works is of the same order of magnitude as that for two
one-photon amplifiers, and so there is no significant disadvantage to
using the higher order device. For higher numbers of one-photon amplifiers
the advantage is greater as the probability that they all work
successfully is so small.

We have quantified the overall device performance using a figure of merit which takes account of both fidelity and success probability. This figure of merit displays the significant advantage to be gained from using the two-photon amplifier.

Furthermore the resources required by the two-photon amplifier: two
photons, one beam splitter, one lossy beam splitter (or lossless for the
sign shift version) and two photon number discriminating detectors, are
smaller than those required for two one-photon amplifiers: two photons,
six beam splitters (including a two-arm interferometer) and four
detectors. This is still true even if the lossy beam splitter is replaced
by a tritter \cite{multiport,tritter} and the discriminating detectors are replaced by a beam
splitter and two standard detectors. The resources required by three or
more one-photon amplifiers, together with the fact that the success
probabilities are so small, render them of limited use when compared to the two
photon amplifier.

The usefulness of the two-photon amplifier is further underlined by what happens if it fails to work. Photons are typically produced by downconversion sources in pairs, and so single photons are typically heralded.  If, for example, only one heralded photon enters BS1 the two-photon amplifier will sometimes operate as a one-photon amplifier. Thus, even when the device fails to work fully it can sometimes produce a state of some use, albeit of smaller overlap with the desired amplified state.

The extension of nondeterministic amplification to higher basis sets such as those which include three or four photons is possible, but not for the relatively simple scissors device considered in this paper. There are not enough free parameters in the set of beam splitter coefficients to allow the required gain conditions to be satisfied. It ought to be possible to use the same basic scissors structure to perform amplification in such higher bases, but with the beam splitters which make up the device substituted with higher order multiports, with multiple photon inputs. The resources required for higher order multiports are lower than would be used by multiple single photon scissors, so there may be be some advantage to this.

\section*{Appendix: Phase conditions for lossy beam splitters}

The phase condition for a lossy beam splitter is easily derived by first assuming that coherent states are incident from each input arm, and requiring that the mean number of photons in the output does not exceed that of the input. A similar analysis was used in \cite{barnettlossbs} to derive magnitude restrictions. For example if coherent states $|\alpha \rangle$ and $|\beta \rangle$ are incident from ports 1 and 2 on to BS1 this means that (dropping the subscripts)
\bea
|t^\prime \beta + r \alpha|^2 + |t \alpha +r^\prime \beta|^2 \leq |\alpha|^2 +|\beta|^2.
\eea
This can be rearranged to give
\bea
\nn &&\frac{\alpha}{\beta}\left( t^\prime r^* + t^* r^\prime \right) + \frac{\alpha^*}{\beta^*}\left( t^{\prime *} r + t r^{\prime *} \right) \\
&\leq& \left( 1 +\frac{|\alpha|^2}{|\beta|^2}\right) \left( 1-|t|^2 -|r|^2 \right),
\eea
and then
\bea
\nn && \frac{2|\alpha/\beta|}{1+|\alpha/\beta|^2} \frac{2|t||r|}{1-|t|^2 - |r|^2} \cos \left[ (\Phi_r - \Phi_t)/2 \right]\\
&\times& \cos\left[ \phi_\alpha -\phi_\beta +(\phi_r-\phi_r^\prime +\phi_t - \phi_t^\prime)/2 \right] \leq 1.
\eea
where we have explicitly written the coherent state amplitudes in terms of magnitude and phase. The maximum value of the first factor is 1 for coherent states of the same magnitude. We can also adjust the phase of the second cosine factor using the coherent states so that it is $\pm$1. Equation (\ref{rtcondition}) imposes $|r|=|t|$, and so 
\bea
\label{lossphasecond}
\pm \frac{2|t|^2}{1-2|t|^2} \cos \left[ (\Phi_r - \Phi_t)/2 \right] \leq 1.
\eea
For the 50/50 beam splitter this condition is equivalent to Eq. (\ref{phasecond}), and the phase $\Phi_r - \Phi_t$ is limited to $\pm \pi$. For a lossy beam splitter a range of phases around these values is allowed, which increases as the loss grows. Clearly, when the loss reaches 50\% ($|t|^2=|r|^2=0.25$) any phase is allowed. 

For the amplifier we require BS1 phases to be related by 
\bea
\Phi_r - \Phi_t = \pm \frac{2 \pi}{3}.
\eea
If the allowed range of phases around $\pm \pi$ is to include this then Eq. (\ref{lossphasecond}) implies that the transmission coefficient should be limited to $|t|^2 \leq 1/3$.

\end{document}